\documentclass[pra,twocolumn]{revtex4-1}   	

\usepackage{geometry}                		
\usepackage{graphicx}				
\usepackage{amsmath}
\usepackage{amssymb}
\usepackage{color}
\usepackage[ruled,linesnumbered]{algorithm2e}
\usepackage{algpseudocode}

\begin{document}
\title{Monte Carlo Graph Search for Quantum Circuit Optimization}

\begin{abstract}
The building blocks of quantum algorithms and software are quantum gates, with the appropriate combination of quantum gates leading to a desired quantum circuit. Deep expert knowledge is necessary to discover effective combinations of quantum gates to achieve a desired quantum algorithm for solving a specific task. This is especially challenging for quantum machine learning and signal processing. For example, it is not trivial to design a quantum Fourier transform from scratch. This work proposes a quantum architecture search algorithm which is based on a Monte Carlo graph search and measures of importance sampling. It is applicable to the optimization of gate order, both for discrete gates, as well as gates containing continuous variables. Several numerical experiments demonstrate the applicability of the proposed method for the automatic discovery of quantum circuits.
\end{abstract}

\author{Bodo Rosenhahn}
\address{Institute for Information Processing (tnt/L3S), Leibniz Universit\"at Hannover, Germany}

\author{Tobias J.\ Osborne}
\address{Institute of Theoretical Physics and L3S, Leibniz Universit\"at Hannover, Appelstrasse 2, 30167 Hannover, Germany}

\date{\today}			
\maketitle

Quantum computing has brought about a paradigm shift in information processing and promises breakthroughs in the solution of industrial use cases \cite{ls_leimeister,preskill2018quantum,endo2018practical}, physics \cite{jordan2012quantum,bargassa2021quantum}, medicine \cite{Davids2022}, chemistry \cite{cao2019quantum}, biology \cite{marx2021biology}, robotics \cite{mannone2023modeling}, general pattern recognition \cite{bapst2020pattern}, machine learning \cite{mott2017solving,wu2021application,WILLSCH2020107006}, and much more. These opportunities are ameliorated, however, by the significant challenges arising in the discovery and application of quantum software. This is because the design of quantum algorithms still requires expert knowledge. Further, since quantum devices are still small and costly, it is essential to optimize resources, in particular, gate counts. Thus the discovery and optimization of quantum circuits is of significant near-term relevance.

Methods for the automated search for optimal quantum circuits have been investigated in the literature, and the term \emph{Quantum Architecture Search} (QAS) has been adopted to describe this body of research. The name is borrowed and adapted from \emph{Neural Architecture Search} (NAS) \cite{Miikkulainen2020,xie2018snas}, which is devoted to the study and hyperparameter tuning of neural networks. Recent works on QAS are often specific to a problem setup, e.g., it has been applied to quantum circuit structure learning \cite{Ostaszewski2021structure}
 for finding the ground states of Lithium Hydride
and the Heisenberg model in simulation, as well as for finding the ground state of a Hydrogen
gas. Many QAS-variants are focussed on discrete optimization and exploit optimization strategies for non-differentiable optimization criteria. Here
variants of Gibbs sampling 
\cite{PhysRevResearchLi20}, evolutional approaches  \cite{9870269}, genetic algorithms \cite{RasconiOddi2019,PhysRevLett116.230504}, neural-network based predictors \cite{Zhang2021}, variants with noise-aware circuit learning \cite{PRXQuantum21}, and the optimization of approximate solutions \cite{PhysRevX10021067} have been suggested. A recent survey on QAS can be found in \cite{Zhu23ICACS}. Going beyond discrete optimisation it is also possible to exploit gradient-descent based optimization schemes \cite{quantum3020021,Zhang2022} or reinforcement learning \cite{PhysRevResearch.2.033446} for QAS.

The recent work \cite{WangAQC23} proposes a Monte Carlo Tree Search (MCTS) based on a multi-armed bandit formulation. This paper is closest in spirit to our present work, and the promising results and challenges identified there inform and inspire our investigation. In particular, since a tree cannot contain cycles, the rollout process (sometimes called \textit{simulation}), can generate nodes and branches which are already part of the tree, leading to multiple identical circuits in the search. This is a consequence of the locality of quantum gate sets, with parallel operations creating cycles (see Fig.~\ref{fig:QGraph} for an illustration).  

In this paper we overcome the challenges presented by cycles in the quantum circuit graph by proposing the use of Monte Carlo Graph search (MCGS) \cite{8632344} to optimize a combination of mixed discrete and continuous variables. This is possible since most gates containing continuous variables are smooth (e.g.\  consisting of rotation coefficients) and thus it is possible to automatically extract the Jacobians for a fast gradient descent while optimizing the quantum computation graph. Our contributions can be summarized as follows:
\begin{enumerate}
\item We propose a Monte Carlo graph search algorithm for quantum architecture optimization.
\item Our model allows for a joint discrete and continuous optimization of the quantum gate ordering and parameters.
\item Several applications demonstrate its applicability, e.g., for the optimization of the quantum Fourier transform, diverse quantum cellular automata, and simple quantum machine learning tasks.
\item Our source code for optimization will be made publicly available.
\end{enumerate}

\section{Preliminaries}
In this section we give a brief overview of the physical systems we discuss in the sequel, and provide a description of the architecture search optimization strategies that we compare and contrast. In particular, reference methods frequently used for discrete optimization are briefly introduced. They are later used for a direct comparison with our proposed MCGS algorithm.

We focus on the setting where our quantum information processing device is comprised of a set of $N$ \emph{logical} qubits, arranged as a quantum register (see, e.g., \cite{10.5555/1206629,nielsenQuantumComputationQuantum2000} for further details). Thus the Hilbert space of our system is furnished by $\mathcal{H}\equiv (\mathbb{C}^2)^{\otimes N} \cong \mathbb{C}^{2^N}$. In this way, e.g., a quantum state vector of a 5-qubit register is a unit vector in $\mathbb{C}^{32}$. We assume throughout that the system is not subject to decoherence and remains pure. (The extension of the results presented here to the mixed-state case will be the subject of a future investigation.)

Quantum gates are the basic building blocks of quantum circuits, similar to logic gates in digital circuits \cite{Selinger2004a}. According to the axioms of quantum mechanics, quantum logic gates are represented by unitary matrices so that a gate acting on $N$ qubits is represented by a $2^{N}\times 2^{N}$ unitary matrix, and the set of all such gates together with the group operation of matrix multiplication furnishes the symmetry group $\textsl{U}(2^N)$. In order to describe explicit matrix representations we exploit the \emph{computational basis} $\{|x_1,x_2,\ldots, x_N\rangle\,|\, x_j \in \{0,1\}, j = 1, 2, \ldots, N\}$ furnished by the eigenstates of the Pauli $Z$ operator on each qubit $j$.

Standard quantum gates include the Pauli-($X$, $Y$, $Z$) operations, as well as Hadamard-, $\textsc{cnot}$-, $\textsc{swap}$-, phase-shift-, and $\textsc{toffoli}$-gates, all of which are expressible as standardised unitary matrices with respect to the computational basis. The action of a quantum gate is extended to a register of any size exploiting the tensor product operation in the standard way. Most gates do not involve additional variables, however, e.g., a phase-shift gate $R_X(\theta)$ applies a complex rotation and involves the rotation angle $\theta$ as free parameter. This parameter should then be jointly optimized together with the architecture of an overall quantum circuit.

A quantum circuit of length $L$ is then described by an ordered tuple $(O(1), O(2), \ldots, O(L))$ of quantum gates; the resulting unitary operation $U$ implemented by the circuit is the product 
\begin{equation}
  U = O(L)O(L-1)\cdots O(1).  
\end{equation}  

\subsection{Genetic algorithms}
A Genetic Algorithm (GA) belongs to the family of so-called evoluationary algorithms. A GA is a population-based metaheuristic  inspired by biological evolution.
It comprises a fitness function to evaluate individuals of a  population, a selection process (driven by the fitness scores) to decide which individuals are used for reproduction and genetic operators, such as crossovers, and mutations to generate new individuals. These new individuals form a new generation which is further evaluated in an ongoing evolution. Genetic algorithms are commonly exploited in discrete optimization and the interested reader is referred to \cite{rug01Eiben} for further details.

For the numerical experiments carried out in this paper the fitness function is directly given by the optimization task (a loss or quality score). Each individual (quantum circuit) $I_1$ is represented by an ordered tuple of quantum gates, e.g., $I_1=(O_1(1), \ldots, O_1(n))$. 
For a crossover between two circuits $I_1$ and $I_2$, a point on both parents' chromosomes is randomly picked which is called the  \textit{crossover point}. Gates to the right of that point are swapped between the two parent chromosomes. This results in two children, $$(O_1(1), \ldots, O_1(j), O_2(j+1), \ldots,  O_2(n))$$ and $$(O_2(1), \ldots, O_2(j), O_1(j+1), \ldots,  O_1(n)),$$ each carrying some genetic information from both parents. A mutation is then furnished by a random exchange of a quantum gate. A recent work on evolutionary quantum architecture search for parametrized quantum circuits was presented in \cite{DingGecco22}.
  
\subsection{Particle filter}
Particle filtering uses a set of samples (which are then called particles) to model a posterior distribution of a stochastic process given some observations. A particle filter is also called a \emph{sequential Monte Carlo method} \cite{wills2023sequential}. These are Monte Carlo algorithms which are commonly used to find approximate solutions for filtering problems of nonlinear state-space systems. More recently they have been applied to quantum systems as Quantum Monte Carlo methods \cite{gubernatis_kwerner_2016}. In the controls literature, particle filters are exploited to estimate the posterior distribution of the state $x_t$ of a dynamical system at time $t$ conditioned on the data,
$p(x_t|z^t,u^t)$. This posterior is estimated via the following recursive formula
\begin{widetext}
\begin{eqnarray}
p(x_t|z^t,u^t) &=& \eta_t\; p(z_t|x_t)\int p(x_t|u_t,x_{t-1})\;
p(x_{t-1}|z^{t-1},u^{t-1})\;dx_{t-1}, 
\label{eqBF}
\end{eqnarray}
\end{widetext}
where $\eta_t$ is a normalization constant. 

Three probability distributions are required for such a particle filter: (1) a so-called {\em measurement model}, $p(z_t|x_t)$, which gives the probability of measuring $z_t$ when the system is in
state $x_t$; (2) a {\em control model}, $p(x_t|u_t,x_{t-1})$, which models the effect of a control $u_t$ on the system state. It provides the probability that the system is in state $x_t$ after
executing control $u_t$ at state $x_{t-1}$; and (3) an {\em initial state
distribution} $p(x_0)$ is required, to specify the user's knowledge about
the initial system state, see also \cite{NIPS2001_Thrun}. In computer vision, the so-called \emph{condensation algorithm} is a well-known example of how to perform a conditional density propagation for visual tracking \cite{NIPS1996_0829424f}.

The implementation of a particle filter can be very similar to that of a genetic algorithm. It can be based on $M$ independent random variables $\xi_0^i$, $(i=1, \ldots, M)$ with a probability density $p(x_0)$. Based on the underlying distribution, e.g., representing a fitness score, $M$ of these variables are selected $\xi_k^i \rightarrow \hat{\xi}_k^i$ and diffused using a mutation-like operation, yielding a new set $\xi_{k+1}^i$.

\subsection{Simulated annealing}
Simulated Annealing (SA) is another probabilistic technique to approximating the optimum of a given function \cite{Kirkpatrick1983}. The name derives from annealing in metallurgy where the process involves heating and a controlled cooling of a material to change and control its physical properties. 
As an optimization scheme the algorithm works iteratively with respect to time $t$ given a state $x_t$. 
At each step, the simulated annealing heuristic samples a neighboring state $\hat{x}_t$ of the current state $x_t$. Then a probabilistic decision is made to decide whether to move to the new state $x_{t+1}=\hat{x}_t$ or to remain in the former state $x_{t+1}=x_t$.
The probability of making the transition from the current state $x_t$ to the new state $\hat{x}_t$ 
is defined by an acceptance probability function 
$P(e(x_t), e(\hat{x}_t), T)$. The function $e(x)$ evaluates the energy of this state, which is in our case 
the fitness score given by the optimization task (e.g. the $\ell_2$-loss). The parameter $T$ is a time-dependent variable dictating the behavior of the stochastic process according to a cooling scheme or annealing schedule. The $P$ function is typically chosen in such a way that the probability of accepting an uphill move decreases with time and it decreases as the difference $e(\hat{x}_t)-e(x_t)$ increases.
Thus, a small increase in error is likely to be accepted so that local minima can be avoided, whereas a larger error increase is not likely to be accepted. A typical function for $P$ takes the form
\begin{equation}
P(e(x_t), e(\hat{x}_t), T) \propto \exp \left(  
-\frac{e(\hat{x}_t)-e(x_t)}{k T}\right),
\end{equation}
with $k$ a \emph{damping factor} $k>0$.

\subsection{Monte Carlo tree search}
Monte-Carlo tree search (MCTS) is a heuristic search algorithm for decision processes \cite{8632344}. It makes use of random sampling and very efficiently balances the well-known exploration-exploitation dilemma in large search spaces. A typical example is provided by game states where non-promising game configurations are avoided, e.g., typical for board games such as \textit{chess} or \textit{tic-tac-toe}. MCTS is a common approach in reinforcement learning, typically in combination with deep reinforcement learning \cite{Silver2016}. As it visits more interesting nodes more frequently, it grows asymmetrically and focusses the search time on more relevant parts of the tree. 

Saffidine et al.\ \cite{SAFFIDINE201226} present a framework for testing various algorithms that deal with transpositions in MCTS. They call this framework Upper Confidence bound for Direct acyclic graphs (UCD) and apply this formalism to overcome the exploration-exploitation dilemma. Their search strategy in the DAG follows the Upper Confidence bounds for Trees (UCT) algorithm \cite{Auer2002}. These predecessors have been more recently applied to Monte Carlo Graph Search to optimize game play in AlphaZero based reinforcement learning \cite{Czech_Korus_Kersting_2021}.

\section{Probabilistic graphical models}
In this section we describe the graphical model we exploit to characterize the search space of quantum circuits.

We assume throughout that we have a fixed set $\mathcal{OP}=\{O_1, O_2, \ldots\}$ of elementary quantum gates that we are allowed to apply. Note that in $\mathcal{OP}$ the same unitary gate acting on different qubits is considered to be a different elementary gate. E.g., the Pauli-$X$ operator acting on qubit 1, written here as $X(1)$, is considered to be a different elementary gate to $X(2)$, which is the Pauli-$X$ operator acting on qubit 2. Starting with the identity operator $\mathbb{I}$ we can build quantum circuits by selecting elementary gates from $\mathcal{OP}$ and multiplying from the left. (Thanks to the universality theorem \cite{nielsenQuantumComputationQuantum2000} we know that we can approximate an arbitrary unitary to arbitrarily good accuracy with a sufficiently long product of such gates.)

\begin{figure*}
\centering
  \includegraphics[width=0.8\textwidth]{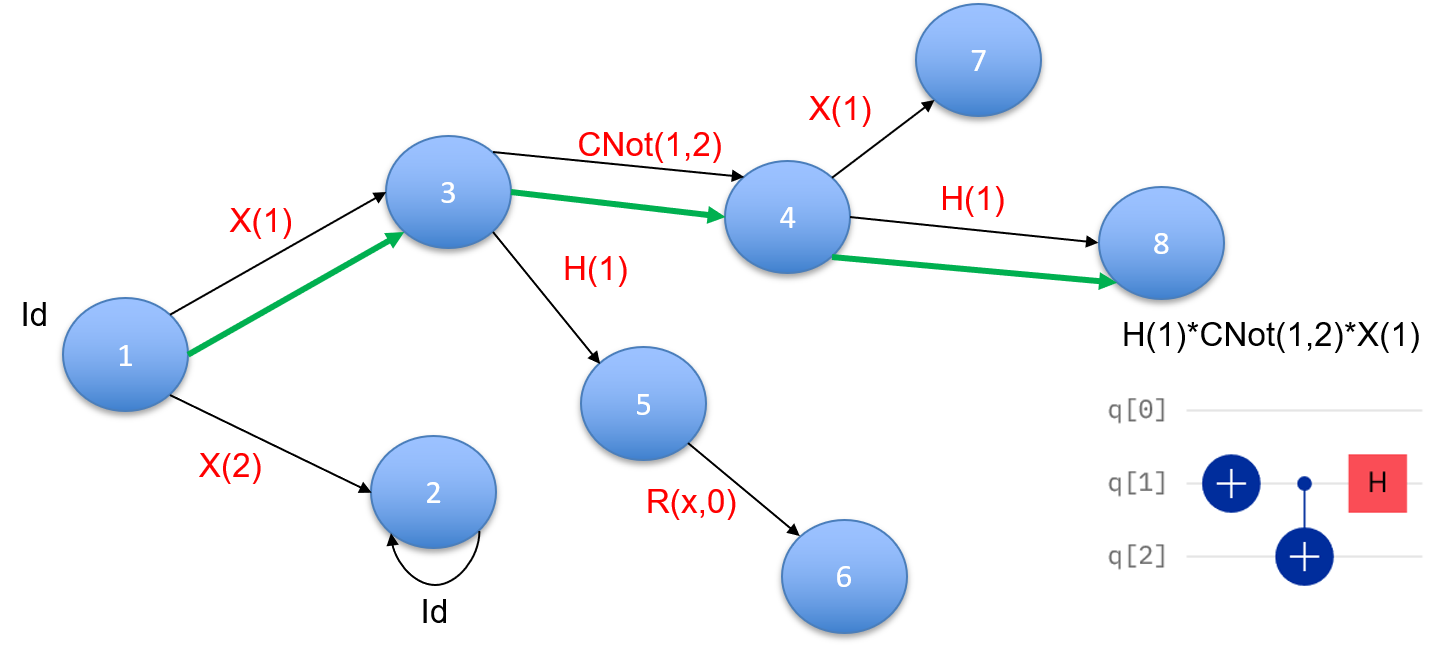}    
\caption{Example graph of quantum circuits. The edges are labelled by possible elementary gates. The vertices are identified with the unitary operator built by taking the product of the gates along the shortest path. A trajectory on this graph directly corresponds to a quantum circuit.
}
\label{fig:QGraphV1}
\end{figure*}

We associate a vertex from a vertex set $V$ to each quantum circuit built from a product of elementary gates from $\mathcal{OP}$. We connect with an edge two such vertices if the corresponding quantum circuit differs (on the left) by an elementary gate. In this way, a collection of quantum circuits is endowed with a graph structure, with vertices decorated by quantum circuits and edges weighted by elementary gates. In Fig.~\ref{fig:QGraphV1} a tiny example graph for differently ordered quantum gates is depicted. The edges are labelled by possible gates of the quantum circuit. The nodes are decorated by the resulting unitary when concatenating the operations along the shortest path. Thus, each node is identified with a possible quantum circuit. It is important to note that this graph contains cycles since identical quantum circuits have multiple representations with different gates and gate orders.

\begin{figure*}
\centering
  \includegraphics[width=0.8\textwidth]{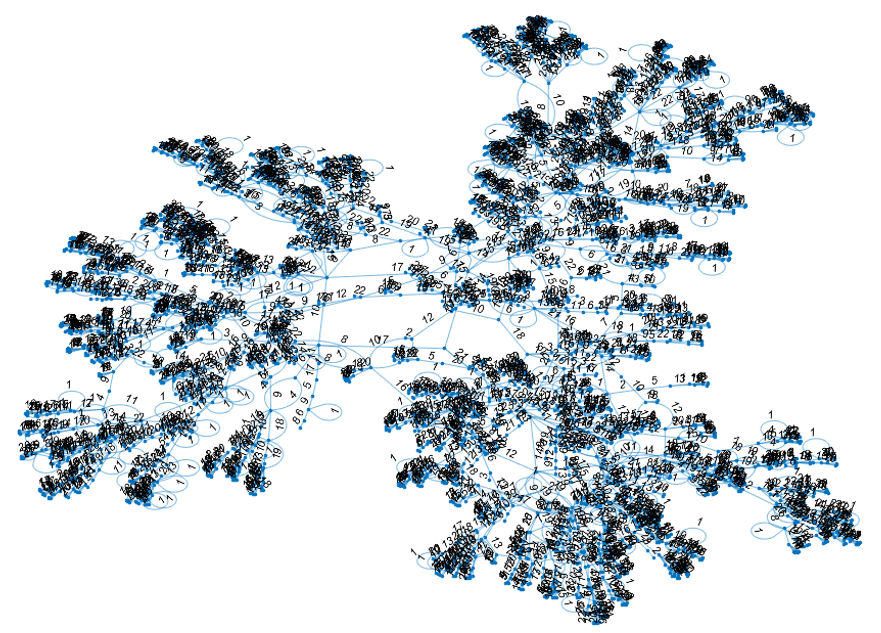}    
\caption{Example graph of quantum circuits
}
\label{fig:QGraph}
\end{figure*}

In Fig.~\ref{fig:QGraph} we illustrate a few steps of such a growing graph model. As the depth of the graph grows exponentially with the number of gates and nodes, it is computationally infeasible to precompute such a graph for all possible circuits. E.g., a tiny set of $20$ elementary quantum gates can be assembled to build $20^{5}$ combinations for quantum circuits of length $5$. Thus it is not possible to evaluate all configurations in a feasible time to solve for a specific optimization task. 

Since the graph model generated by the evaluation of quantum gates can have cycles, we apply a Monte Carlo search on the graph model with quantum gates as transitions. Thus, given a specific task, every node will receive a quality score which is used to compute a probability for the a selection of this node. Based on the random selection and already explored operations, a new operation is randomly selected to grow the graph. Once a solution is found, an efficient quantum circuit can be generated by computing the shortest path in the graph from the start node to the target node.

This strategy is formalised as follows: We have a set of vertices $V$ -- associated with quantum circuits -- of a graph $G=(V;E)$ with $V=\{v_1, \ldots v_n\}$ and we build a probability function $p(v_i)$, $\sum_i p(v_i)=1$, which assigns to each node of the graph a probability for selection. Poisson sampling is then exploited as the underlying sampling process. It is assumed that each vertex of the graph is an independent Bernoulli trial. Following standard mathematical conventions, the first-order inclusion probability of the $i$th element of the graph is denoted by the symbol $\pi_i = p(v_i)$. We further associate to each vertex of the graph an underlying task specific \emph{quality score} $s_i\ge 0$. (Here larger values of $s_i$ imply better quality.) Accordingly, we compute the first-order inclusion probability via 
$\pi_i=\frac{s_i}{\sum_j s_j}$. This paradigm of
Monte Carlo Search 
\cite{doi:10.1080/01621459.1949.10483310} and adapted Gibbs sampling \cite{george1993variable} is used to iteratively grow a graph containing the effects of ordered quantum operations as trajectories in this graph, see Figure 
\ref{fig:QGraphV1}. 

\begin{figure*}
\centering
  \includegraphics[width=\textwidth]{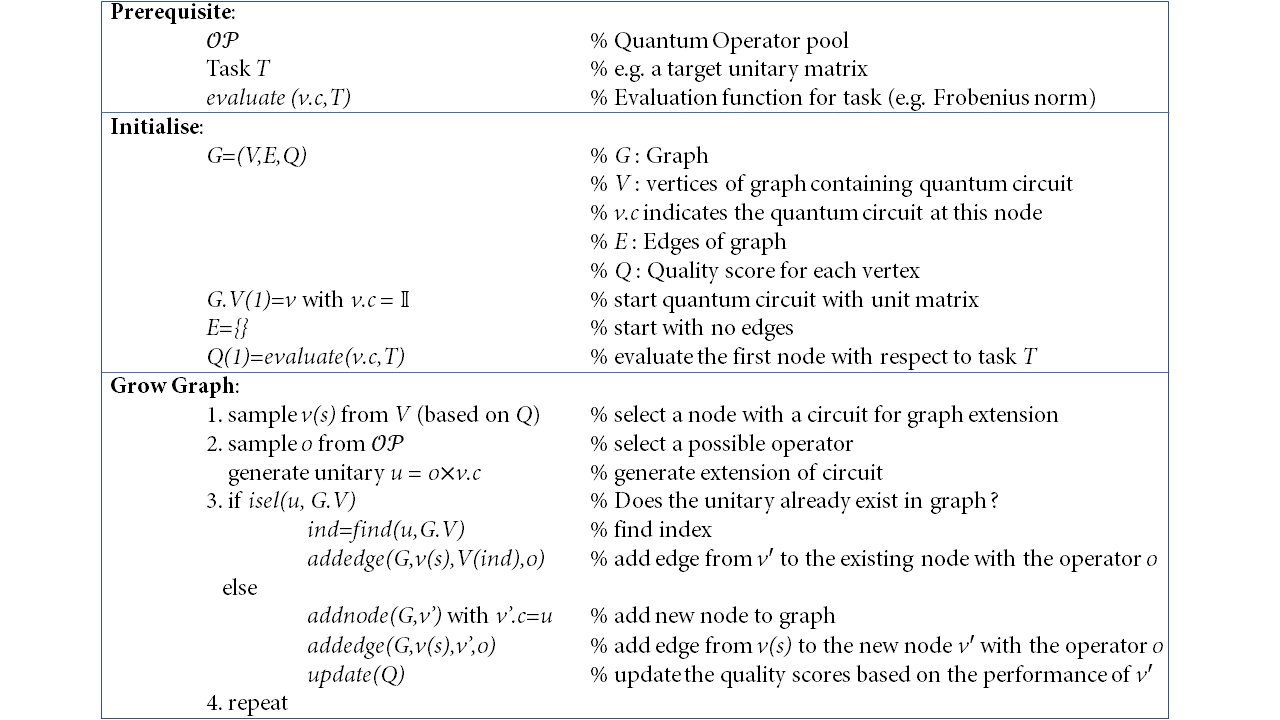}    
\caption{Basic steps of the Quantum MCGS Algorithm}
\label{fig:QCAlgorithm}
\end{figure*}

Our MCGS procedure is now as follows: Given a set of vertices $V=\{v_1, \ldots v_n\}$ and a probability function
$p(v_i)$, Poisson sampling is used to select an existing node $v_i$ from the graph. A quantum gate is then randomly selected from the set $\mathcal{OP}$ of elementary gates according to a uniform distribution and applied  (from the left) to the quantum circuit associated with $v_i$. The result is a new circuit, which is either associated with an existing node or a new one \footnote{At this stage it may be convenient to introduce an approximation factor $\epsilon$: Two circuits $U$ and $V$ are then considered to be equivalent if $||U-V||\le \epsilon$.}. If the circuit at a node already exists, e.g.\ $v_j$, a new edge $(v_i, v_j)$ is added to the graph, decorated with edge weight given by the applied elementary gate. If the quantum circuit does not exist, the graph is extended by adding a new node $v_{N+1}$ and an edge  $(v_i, v_{N+1})$.
Since the probabilities vary with respect to the quality score of the nodes, the graph grows asymmetrically. Once a node is reached which (sufficiently accurately) solves the optimization task, a shortest path from $v_1$ to the target node gives the shortest available quantum circuit, see also Figure \ref{fig:QGraphV1}. Figure \ref{fig:QCAlgorithm} summarizes the basic steps of the MCGS Algorithm.

Please note that such a graph can also be reused for different kinds of optimizations and it can be analyzed very generally to identify cycles, clusters and other structural properties on the effect of quantum gates.  

\subsection{Optimization of Continuous Variables}
Several gates can contain continuous variables for optimization, e.g.\ phase shift gates 
\begin{equation}
P(\phi) = \left( 
\begin{matrix}
1 & 0 \\ 0 & \exp(i \phi)
\end{matrix}
\right) 
\end{equation}
with corresponding Jacobians:
\begin{equation}
\frac {\partial P(\phi)}{\partial \phi} = \left( 
\begin{matrix}
0 \hspace{0.4cm}& 0 \\ 0 \hspace{0.4cm}& i\exp(i \phi)
\end{matrix}
\right)
\end{equation}
Since the involved functions are smooth and differentiable,
the Jacobian of such a matrix and the Jacobian for a chain of operations is easy to compute via the product rule. Thus, given
 a quantum circuit of length $L$ which is described by an ordered tuple $(O(1), O(2), \ldots, O(L))$ of quantum gates and $\Phi=(\phi_1 \ldots \phi_k)$ continuous variables within this chain,
 the resulting unitary operation is then a function 
 $ U(\Phi)=U(\phi_1, \ldots, \phi_k)$.
This function is typically used for the optimization of a loss function, for example the distance from a target matrix $O$, e.g., a DFT matrix. For numerical and implementation convenience the loss function was chosen to be the Frobenius norm between $O$ and $U(\Phi)$,
\begin{multline}
L(\Phi)= \| O - U(\Phi) \|_F = \\ \sqrt{\text{tr}\left[(O-U(\Phi))^\dag(O-U(\Phi))\right] }.
\end{multline}
Although the Frobenius norm does not have a simple operational interpretation, it is easy to compute both numerically and also experimentally via a {\sc swap} test.

The Jacobian of the loss function is given by
\begin{equation}
 \nabla_\Phi  L(\Phi) = 
 \left[ \frac{\partial L(\Phi)}{\partial \phi_1}, \ldots ,\frac{\partial L(\Phi )}{\partial \phi_k}  \right].
\end{equation}
This vector can be numerically computed and used for optimization by using automatic differentiation.
Optimization of the involved parameters can be then carried out with a gradient descent iteration using
$\Phi^{t+1} = \Phi^{t} - \eta \nabla_\Phi L(\Phi)$.
Here, $\eta$ denotes a damping factor which has been set to $0.2$ for all our experiments.


\section{Experiments}

Since the MCGS algorithm we describe here is capable of optimizing both discrete and mixed discrete and continuous settings, the experiments are divided in two parts. In the first part only results for discrete optimizations are presented. In the second part we also describe some experiments involving mixed discrete and continuous variables.

\subsection{Discrete quantum architecture search} 

The first experiment we conducted targeted the optimization of the quantum Fourier transform. It is well known that the steps of the radix-2 FFT can be realised as a quantum circuit using primarily phase shift and Hadamard gates. E.g., for an 8-dimensional FFT, three qubits are sufficient and 7 gates can be used to compute the FFT matrix.

\begin{figure}
\centering
 \includegraphics[width=\columnwidth]{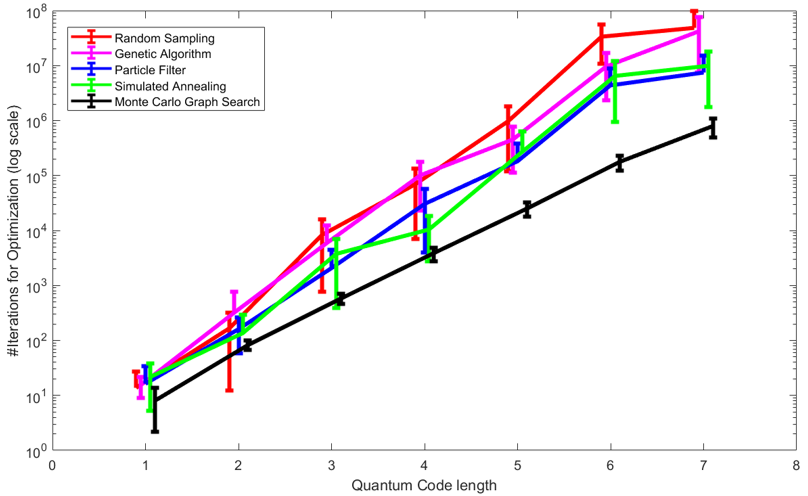}    
\caption{Amount of required samples for quantum circuit optimization for differing circuit lengths. Compared are random sampling, a genetic algorithm, a particle filter, simulated annealing and the proposed MCGS model. The bars depict the standard deviation. The diagram is illustrated with respect to a logarithmic $y$ axis scale. One observes in this way the the MCGS already requires an order of magnitude fewer samples than the four baselines.}
\label{fig:QuantInc}
\end{figure}

In the first experiment we compared the optimization of quantum circuits of length $1$, $2$,  $\ldots$, $7$, for a given predefined set of elementary quantum gates. Note, for a database of elementary gates of size $32$ and a circuit length of $7$ there are $32^7 \sim 3.5 \times 10^{10}$ different quantum circuits possible; the search space grows exponentially and is already infeasibly large (for exhaustive searches) for larger circuits. In Fig~\ref{fig:QuantInc} we present the comparison of four different baselines, based on a naive random sampling, a genetic algorithm (Sec~3.2), a particle filter (Sec.~3.3), and simulated annealing (Sec.~3.4), respectively, along with our proposed Monte Carlo Graph Search. The $x$-axis records the circuit length required for a predefined quantum circuit. The complexity of the optimization problem increasing exponentially with circuit length.
Note, that the $y$ axis is scaled with the log function, so that a linear slope of indicates an exponential growth in complexity. The error bars depict the standard deviation.  For this experiment, the amount of experiments/observations was set to $10$. In comparison to the four baselines, our proposed  Monte Carlo Graph Search requires far fewer samples to reach a solution. One explanation for this is that unnecessary samples (e.g., leading to cycles, etc.) are efficiently avoided.  

\begin{figure}
\centering
 \includegraphics[width=\columnwidth]{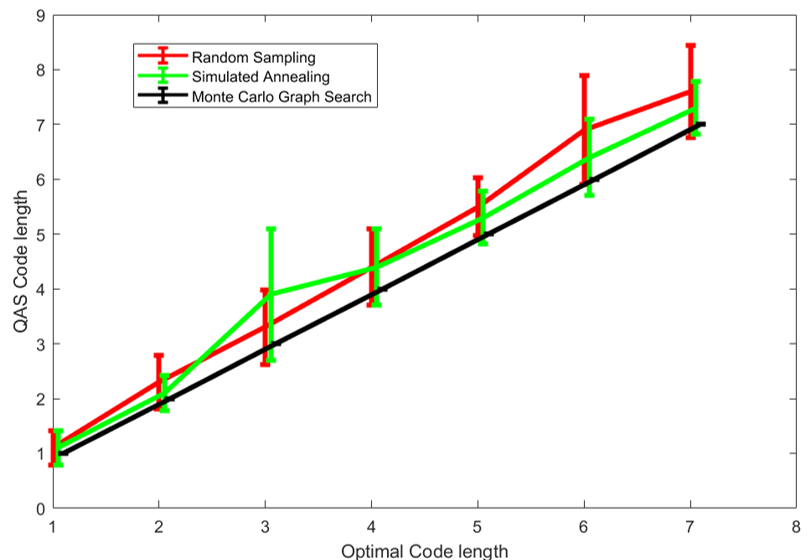}    
\caption{Optimal circuit length versus circuit length of different architecture search algorithms. }
\label{fig:CodeLength}
\end{figure}

In the next experiment we analyze the efficiency of the generated quantum circuits in terms of the amount of required gates. As the computation graph contains cycles, a valid question is, if standard sampling based approaches for QAS can lead to solutions which require more gates than necessary, resulting in inefficient code. 
The approaches of Gibbs-Sampling, simulated annealing, and the proposed MCGS model allow one to optimize quantum circuits where the resulting code length is not fixed. Due to the diffusion and crossover steps, it is not clear how one can do this for the particle filter and the genetic algorithms, so these approaches were omitted for this experiment. In Figure \ref{fig:CodeLength} the mean and standard deviation of the random sampling, simulated annealing, and MCGS for varying optimal circuit length tasks is shown. The $x$-axis is labelled by the optimal circuit length and the $y$-axis by the required circuit lengths (including standard deviation) of the different optimizers. The optimal graph is a straight line, which was achieved by our proposed MCGS. Thus, the MCGS always finds the optimal length, whereas the sampling and annealing schemes more often tend to find inefficient solutions. Thus, the proposed MCGS jointly ensures efficient models during optimization.

\subsection{Quantum circuits for classical cellular automata}
A cellular automaton (CA) is a mapping on a set of states of connected cells (e.g.\ arranged as a graph).
Each cell has a cell state (e.g.\ binary) and which changes according to a predefined set of rules given by the local neighbors of each cell. Simple and nontrivial cellular automata are furnished already for one-dimensional graphs, with two possible states per cell. The neighbors are defined as the direct adjacent cells on either side of the cell. In this setting the rules for changing the state of each cell can be defined as a mapping from a 3-dimensional binary state to a new binary state. Thus, there are $2^3=8$ patterns for a neighborhood. There are then $2^8 = 256$ possible combinations for rules which describe $256$ different cellular automata. In general they are referred to by their Wolfram code and are called \textit{R-X automata}, with $X$ being a number between $0$ and $255$. Several papers have analyzed and compared these 256 cellular automata. The cellular automata defined by rule 30, rule 90, rule 110, and rule 184, are particularly interesting and are given by the mappings

\begin{table}
\begin{tabular}{|c||cccccccc|}
\hline
Rule & 111&110&101&100&011&010&001&000 \\
\hline
30(=00011110)&0&0&0&1&1&1&1&0 \\
90(=01011010)&0&1&0&1&1&0&1&0 \\
110(=01101110)&0&1&1&0&1&1&1&0 \\
184(=10111000)&1&0&1&1&1&0&0&0 \\
\hline
 \end{tabular}
 \end{table} 
 
Rule 30 exhibits so-called class 3 behavior. This means that even simple input patterns lead to chaotic, and seemingly random histories. When starting from a single live cell, Rule 90 produces a spacetime diagram resembling the Sierpinski triangle. Rule 110, similar to \textit{Game of Life}, exhibits what is called class 4 behavior, which means it is neither completely random nor completely repetitive. Finally, Rule 184 is notable for solving the majority problem as well as for its ability to simultaneously describe several, seemingly quite different, particle systems. For examples Rule 184 can be used as a simple model for traffic flow.
Thus, several of these transition rules exhibit interesting aspects for researchers in mathematics and optimization, as well as biology and physics. Due to the simple definition of these transition rules, and their consequent rich behaviour, these CA provide a diverse and interesting class of targets for quantum circuits encoding.

\begin{figure}
\centering
  \includegraphics[height=0.15\textheight]{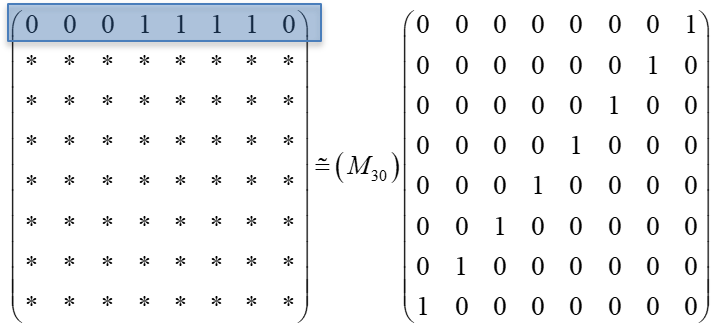} 
    \includegraphics[height=0.15\textheight]{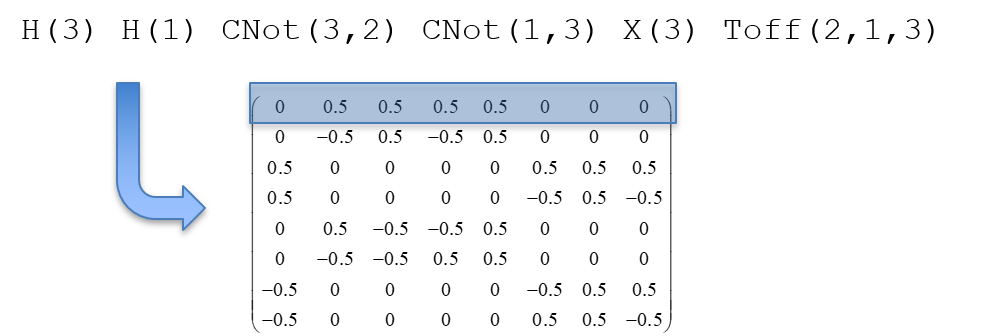}  
\caption{Concept for a classical cellular automaton as a target for a quantum circuit. In this case the $M_{30}$  automaton is targeted: The binary $8$-dimensional input vector encodes the 3D binary state using the one-hot encoding. A quantum circuit is searched for that, when applied to $|000\rangle$ produces the corresponding $8$ dimensional vector according to the rule that if the probability is larger than zero (or a small threshold), the bit is set to 1, else to zero. 
}
\label{fig:QCA1}
\end{figure}
Fig.~\ref{fig:QCA1} visualizes the basic concept.  Given an 8-dimensional one-hot encoded input vector (encoding the 3 input binary elements), the quantum circuit is applied to the initial vector $|000\rangle$. The resulting probability indicates if the new state is zero or one.
 \begin{figure*}
\centering
 \includegraphics[width=0.8\textwidth]{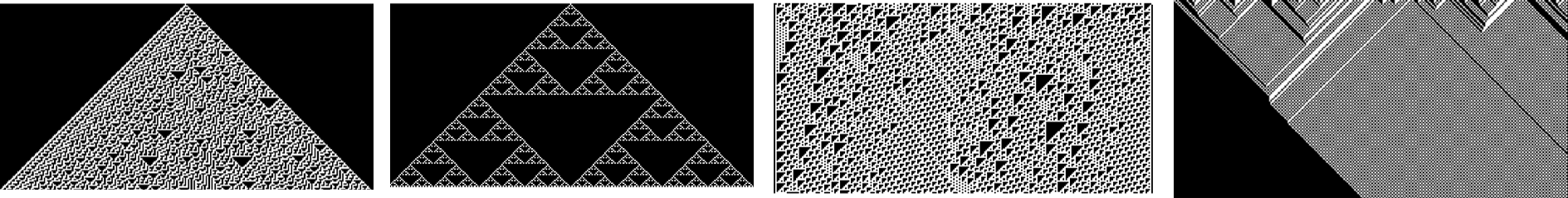}    
\caption{Example realisations of the Cellular automata 30, 90, 110 and 184. The images show the evolution of the pattern over time (from top to bottom).
The two examples at the left side have been initialized with a zero vector and a one at the middle, the examples at the right side have been initialized with a random binary pattern.}
\label{fig:CellAuto}
\end{figure*}

Fig.~\ref{fig:CellAuto} shows example realizations of the cellular automata 30, 90, 110 and 184. The images show the evolution of the pattern over time (from top to bottom). The two examples at the left side of the figure have been initialized with a zero vector and a one entry at the middle, the examples at the right side have been initialized with a random binary pattern.

\begin{figure*}
\centering
  \includegraphics[width=1.0\textwidth]{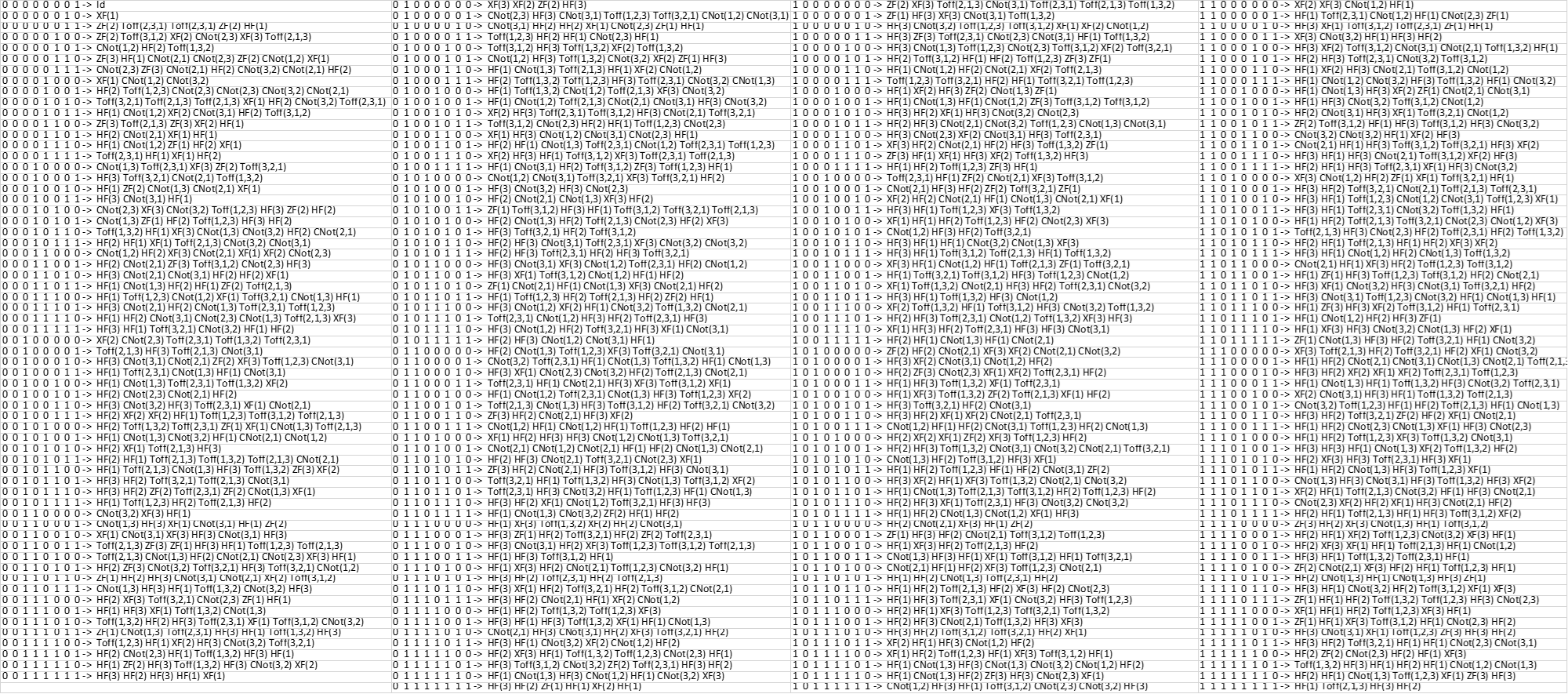}    
\caption{Quantum circuits for vectors encoding RX cellular automata}
\label{fig:QCA}
\end{figure*}

Fig.~\ref{fig:QCA} shows quantum circuits for all possible $8$-entry vectors corresponding to the 256 Wolfram Codes which have been optimized using our quantum architecture search algorithm.

\subsection{Mixed Discrete-Continuous Quantum Architecture Search} 
Our approach exploiting the Monte Carlo Graph Search can also be easily extended to optimize circuits involving a combination of discrete and continuous variables, as outlined before. In the following experiments we use quantum architectures to solve simple machine learning tasks.
%
For the experiments, the classical \textit{wine}, \textit{zoo} and  \textit{iris} datasets were used. The  datasets present multicriterial classification tasks, with three categories for the wine dataset, seven categories for the zoo dataset, and three for the iris dataset.  The datasets are all available at the UCI repository \cite{Dua:2019}.
To model a classification task using a quantum circuit, first the data is encoded as a higher-dimensional binary vector.
Taking the iris dataset as a toy example, it consists of $4$ dimensional data encoding \textit{sepal length}, \textit{sepal width}, \textit{petal length} and \textit{petal width}.
After separating training and test data, a kMeans clustering on each dimension with $k=3$ is used on the training data. Thus, every datapoint can be encoded in a $4
\times 3=12-$dimensional binary vector which contains exactly $4$ non-zero entries. For the given cluster centers, the same can be done with the test data. Thus, a binary encoding is used to represent the datasets.
Table \ref{tab:QML1} summarizes the used datasets, the amount of features (the dimension of each sample), its binary dimensionality, the used qubits to represent the problem as well as the amount of training data, the amount of target classes and the gained accuracy with the optimized quantum model. 
Similar to Fig.~\ref{fig:QCA1}, a part of the quantum register is used to encode the probability of a classification label. The final decision is then based on the highest probability. 

 \begin{figure}
\centering
  \includegraphics[width=\columnwidth]{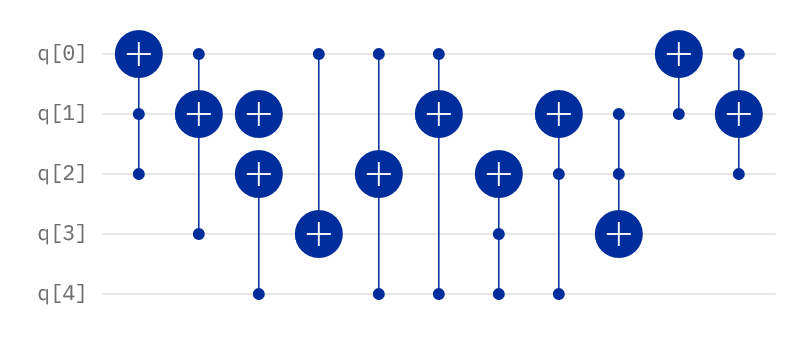}  
 \includegraphics[width=\columnwidth]{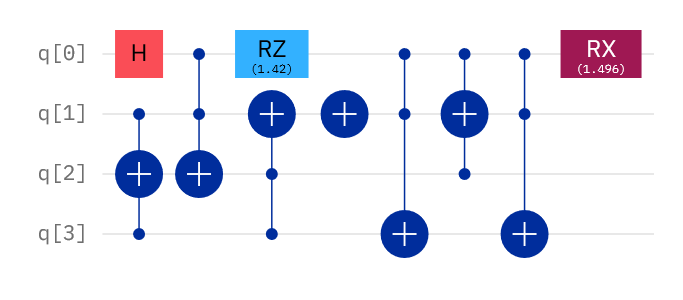}     \includegraphics[width=\columnwidth]{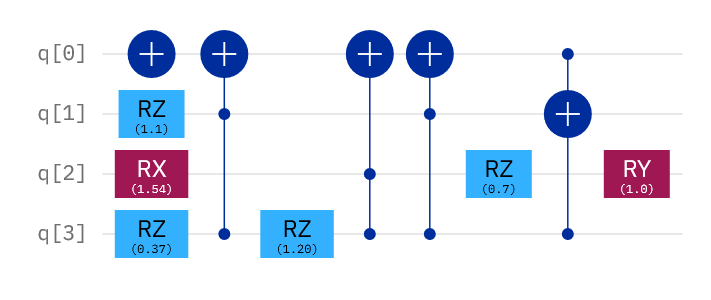}    
\caption{Optimized quantum circuits for the wine, zoo and iris dataset. (Images have been generated using 
{\tt quantum-computing.ibm.com/composer})}
\label{fig:CML1}
\end{figure}

\begin{table}
\centerline{
\begin{tabular}{|c|c|c|c|c|c|c|}
\hline
Dataset & Dim &BDim & qubits& \# Train & \# Classes &Acc  \\
\hline
Iris & 4 & 12 & 4 &100 & 3 & 95 \%  \\ 
Wine & 13 & 26 & 5 & 133 & 3 & 90 \% \\ 
zoo  & 16 & 16 & 4 & 75 & 7 & 92 \%  \\ 
\hline
 \end{tabular}}
 \caption{QML datasets overview and Performance}
 \label{tab:QML1}
 \end{table} 

Figure \ref{fig:CML1} shows example outcomes of the optimized quantum codes for the wine and iris dataset. Note, that the results can vary considerably. This depends on the random selection of training and test data and the random process of the graph generation. Table \ref{tab:QML1} summarizes the three used datasets and the overall performance. Note, that the overall quality is similar to the the results obtained with decision trees or shallow neural networks.

\section{Summary}
In this paper we have proposed a quantum architecture search algorithm based on Monte Carlo graph search and measures of importance sampling. Each trajectory in this graph leads to a quantum circuit which can be evaluated according to whether it achieves a specific task. Our model also allows for the optimization of mixed discrete and continuous gates and several experiments demonstrate the applicability for different tasks, such as matrix factorization, producing cellular automata vectors, and simple machine learning models. A comparison with classical approaches such as greedy sampling, genetic algorithms, particle filter or simulated annealing was carried out and demonstrates that the graph model performs more efficiently, since cycles and redundancies are explicitly avoided. The shortest path from the start node to the target node provides efficient algorithms in terms of circuit length. A future challenge is to overcome the computation time required by the graph model as the number of nodes is increased. One immediate and relevant next step is to generalize the method described here to apply to mixed states and completely positive maps.

\subsection*{Acknowledgments}
This work was supported, in part, by the Quantum Valley Lower Saxony, the Deutsche Forschungsgemeinschaft (DFG, German Research Foundation)–Project-ID 274200144–SFB1227,  under Germanys Excellence Strategies EXC-2123 QuantumFrontiers and  EXC-2122 PhoenixD. 

\end{document}